# Crystal Growth of $Cu_6(Ge,Si)_6O_{18}\cdot 6H_2O$ and Assignment of UV-VIS Spectra in Comparison to Dehydrated Dioptase and Selected Cu(II) Oxo-compounds Including Cuprates.

**Hans Hermann Otto**

Materialwissenschaftliche Kristallographie, Clausthal University of Technology, D-38678 Clausthal-Zellerfeld, Adolph-Roemer-Strasse 2A, Lower Saxony, Germany
* Correspondence: hhermann.otto@web.de

**Abstract:** It is reported on growth of mm-sized single-crystals of the low-dimensional $S$ = ½ spin compound $Cu_6(Ge,Si)_6O_{18}\cdot 6H_2O$ by a diffusion technique in aqueous solution. A route to form Si-rich crystals down to possibly dioptase, the pure silicate, is discussed. Further, the assignment of *dd* excitations from *UV-VIS* spectra of the hexahydrate and the fully dehydrated compound is proposed in comparison to dioptase and selected Cu(II) oxo-compounds using bond strength considerations. Non-doped cuprates as layer compounds show higher excitation energies than the title compound. However, when the antiferromagnetic interaction energy as $J_z\cdot \ln(2)$ is taken into account for cuprates, a single linear relationship between the $D$q$_e$ excitation energy and equatorial Cu(II)-O bond strength is confirmed for all compounds. A linear representation is also confirmed between $^2A_{1g}$ energies and a function of axial and equatorial Cu-O bond distances, when auxiliary axial bonds are used for four-coordinated compounds. The quotient $D$t/$D$s of experimental orbital energies deviating from the general trend to smaller values indicates the existence of $H_2O$ respectively $Cl^{1-}$ axial ligands in comparison to oxo-ligands, whereas larger $D$t/$D$q$_e$ values indicate missing axial bonds. The quotient of the excitation energy $^2A_{1g}$ by $2\cdot{}^2E_g$-$^2B_{2g}$ allows to check for correctness of the assignment and to distinguish between axial oxo-ligands and others like $H_2O$ or $Cl^{1-}$. Some assignments previously reported were corrected.

---

## 1. Introduction

Low-dimensional quantum spin systems are of considerable theoretical and experimental interest together with some applications to which they may lead. In spite of the ability of the $d^9$ transition metal ion $Cu^{2+}$ to form, apart from 3*D* networks, chains, ladders and small clusters, copper compounds are among the most interesting phases. With equal electronegativity compared to silicon, but in contrast to its tetrahedral networks, Cu(II) mainly forms oxo-compounds with chains and networks of connected 'octahedra'.

For instance, copper polygermanate, $CuGeO_3$, has a rather simple crystal structure of 'einer' single chains of $GeO_4$ tetrahedra alongside $S$ = 1/2 spin single chains of edge-sharing $CuO_{4+2}$ octahedra [1] [2]. It was the unique inorganic compound showing the *Spin-Peierls*-transition [3,4]. As a quasi-one-dimensional system it has been the subject of an intensive experimental and theoretical work for the past years. It was a great surprise, when *Otto and Meibohm* [5] succeeded in the synthesis of pure copper polysilicate, $CuSiO_3$, by thermal decomposition of the mineral dioptase, $Cu_6Si_6O_{18}\cdot 6H_2O$. $CuSiO_3$ represents the example of a fully stretched silicate chain structure. It is isotypic to $CuGeO_3$, but does not show the spin-Peierls transition, instead an antiferromagnetic ordering below $T_N$ = 7.9 K [6,7].

The rhombohedral title compound $Cu_6(Ge,Si)_6O_{18}\cdot 6H_2O$ represents a hexacyclo-germanate (silicate) that contains copper-oxygen spiral chains along the *c*-axis, which are connected (intra-chain) by edge-sharing dimers (Figure 1). This structure is interesting because it allows for a quantum phase transition between an anti-ferromagnetically ordered state and a quantum spin liquid [8]. Large quantum fluctuations in green dioptase have been described [9]. Recently, also the germanate analogue, $Cu_6Ge_6O_{18}\cdot 6H_2O$ [10], has been the object of detailed magnetic and structural investigations [11,12].

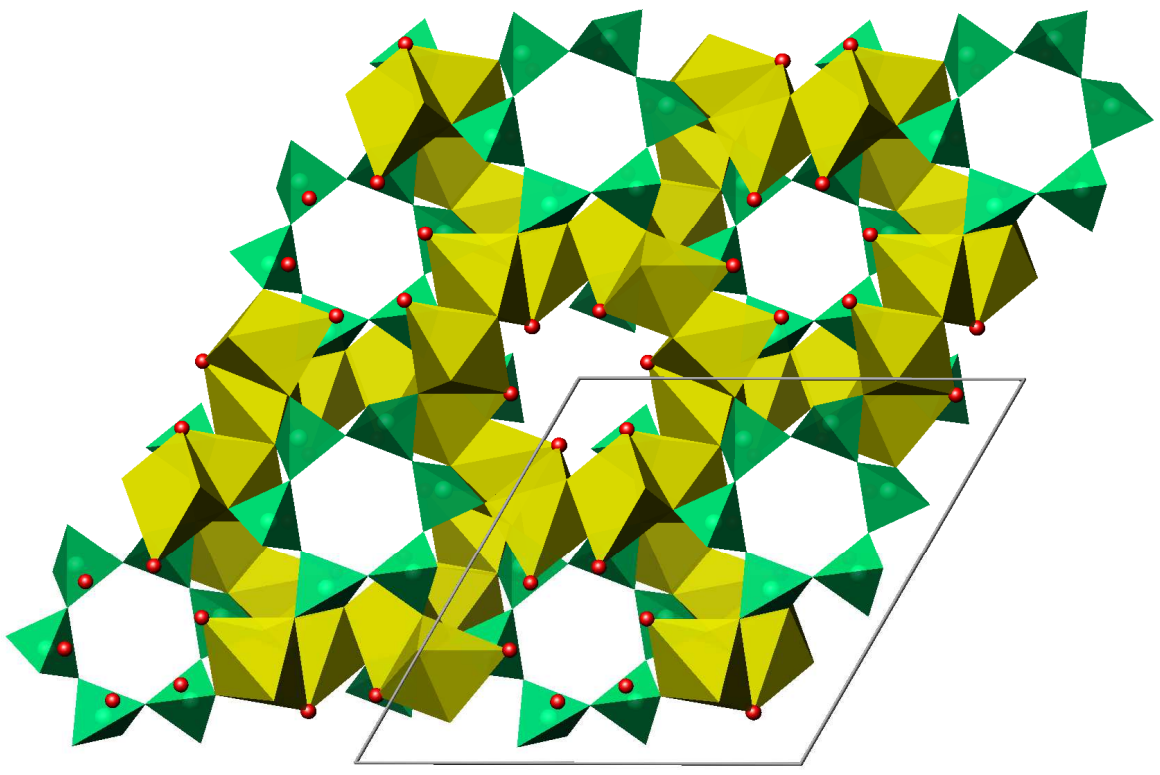

**Figure 1**. Crystal structure of dioptase projected down [001]. A framework of copper oxide octahedra (yellow) screws around the *c*-axis with non-bonding axial water ligands (red) pointing towards empty channels. Six-membered silicate single rings are depicted in green.

If near the empty structural channels located water molecules are removed, a screwed framework of edge-sharing disphenoids rather than flat $CuO_4$ plaquettes remains in the dehydrated compound.

As part of a systematic study of transition metal germanates, silicates and arsenates we have undertaken syntheses of rare copper minerals and new copper compounds in view of its power as low dimensional $S$ = 1/2 spin compounds allowing for interesting physical and physico-chemical properties. First, the synthesis serves not to waste rare mineral specimens for research. There is also the possibility to study an improvement in the crystal growth by replacement of copper by other elements, apart from the chance of doping with electronically or magnetically interesting ones. For example, the replacement of copper by manganese was observed in natural samples of dioptase by *EPR* measurements [13,14].

## 2. Experimental

### 2.1 Crystal Growth

The method described below was used by the author many years earlier for the synthesis of rare minerals, for instance the synthesis of $Pb_3Ge(OH)_6(SO_4)_2\cdot 3H_2O$, the piezoelectric *Tsumeb* mineral fleischerite [15]. For the synthesis of the title compound freshly precipitated gels of $GeO_2$ and $Cu(OH)_2$ were separately filled in 200 ml beaker glasses and thoroughly filled up with distilled water. Then an U-shaped glass pipe of 6 mm inner diameter, well annealed before use to reduce crystal nucleation frequency, was filled free of air bubbles with distilled water. This pipe is then used to connect the distinct solutions in the beakers. Finally, the water surface in the beakers is covered with a film of liquid paraffin to prevent water evaporation and entry of $CO_2$, respectively.
The desired slow diffusion of the distinct solutions into one another leads to the formation of $Cu_6(Ge,Si)_6O_{18}\cdot 6H_2O$ seeds that grow up to 1 mm size of light blue crystals within 8 weeks. Interestingly, most individual crystals form double-crystals. The symmetry situation of this finding must be investigated further. The crystals of stocky prismatic, nearly spherical habit developed {110} and {021} forms (Figure 2).

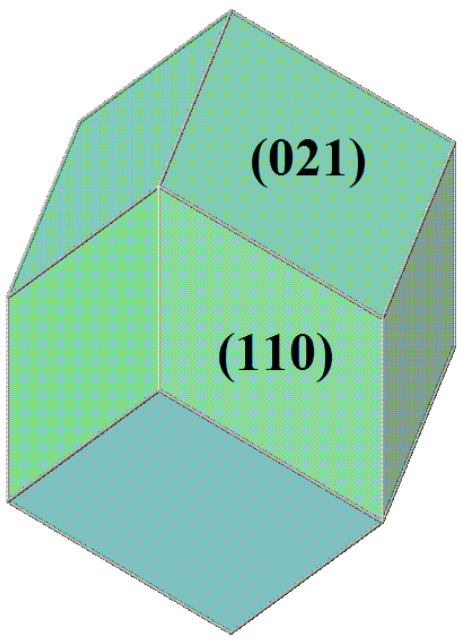

**Figure 2**. Stocky prismatic habitus of the as-grown $Cu_6(Ge,Si)_6O_{18}\cdot 6H_2O$ crystals, showing a combination of the {110} prism and the {021} rhombohedron.

One can extrapolate the time scale to get a crystal of about 2mm diameter and calculate about 1 year of growing time. Trying to exchange Ge by Si by this method seems to be less efficient, only a slightly greenish sheen shows that a small exchange occurred.
The other method of co-precipitation of $GeO_2$, $SiO_2$ and $Cu(OH)_2$ gel and longer time vigorous stirring resulted in a vivid green colored polycrystalline material of about 12 at-% Si determined from lattice parameter changes [13,16]. Also the substitution of some $B^{3+}$ for $Ge^{4+}$ is possible, leading to a beautiful green color [16]. Stirring a longer period and in addition changing the *pH* to more acidic milieu gives at least about 15 at-% Si ($a$ = 14.640 Å, $c$ = 7.806 Å, this work). The effect is based on the different solubility of the Ge-compound in comparison to dioptase. $Cu_6Ge_6O_{18}\cdot 6H_2O$ is easily decomposed by dilute acetic acid, but dioptase does not dissolve. Recently we observed a deepening of color to dioptase green, when the Si-rich solution was exposed to ultrasonic waves, in this way superseding vigorous stirring. The energy that is released when voids implodes (super-cavitation) may be able to assemble more easily and faster the six-membered silicate rings within the cuprate framework of dioptase.
A proposed approach for a possible synthesis of pure polycrystalline dioptase results as follows. The first step will be the spontaneous formation of pure germanate and exchange of maximum Ge by Si through stirring or sonochemical treatment. Then *pH* as well as temperature is altered to increase the solubility of the still Ge-rich compound combined with a simultaneous offer of more Si to form a dioptase layer. A new core of silico-germanate can be grown epitaxially and subsequently transformed to dioptase. Repetition of this process may finally form pure dioptase in mm-sized crystals. An automated process would make sense. Nature has similar tools in the quiver such as rhythmic property changes (concentration, pH, temperature) of metal bearing ascending or descending solutions, apart from a lot of time.

## 2.2 UV-VIS Spectroscopic Investigation

First results of *UV-VIS* spectroscopy on $Cu_6(Ge,Si)_6O_{18}\cdot 6H_2O$ are given in the doctoral theses of my coworkers *Brandt* [17] and *Meibohm* [13], respectively, whereas dioptase itself has been investigated earlier by different researchers [18-21].

*Brandt* [17] reported a color change from turquoise-green to blue on dehydration of dioptase-type copper germanate. In addition, the dehydrated compound showed thermochromic behavior on heating up to 500 °C with a reversible color change to vivid green similar to that of annealed $CuGeO_3$. The color persists when $Cu_6Ge_6O_{18}$ is rapidly cooled down to room temperature. A possible interpretation for this effect is according to [17] the low relaxation rate of the four oxygen ligands around copper. Remember that the equatorial coordination in dioptase is not planar but disphenoidic, and a change to a stiffer, more tetrahedral one may occur with raising temperature.

A reinvestigation of the fully hydrated and dehydrated compounds is primarily undertaken in order to deconvolute and understand the broad UV-VIS spectrum of the synthetic color pigment litidionite, $KNaCuSi_4O_{10}$ [22,23], which shows similarity to that of dioptase.

The room temperature UV-VIS spectra of the samples were taken with the double beam light scanning UV-2501PC CE spectrometer from *Shimazu* with selectable light sources (50W halogen lamp and D2 lamp). The powder sample was coated on a polished aluminium disk and measured in the reflection modus against a $BaSO_4$ standard in the wavelength range between 190 and 900 nm with a spectral bandwidth of 0.1 nm using a 50 nm/min scan and choosing 0.5 nm intervals. From the less structured absorbance profile, recalculated from the measured reflectance, the energy bands were fitted with *Gaussian* profile functions. The better resolved spectra of the dehydrated compounds were fitted first and then the results used as start parameters for the broad spectra of the hydrated compounds.

### 2.3 EPR Data

Electron paramagnetic resonance spectroscopy (*EPR*) provides information about the electronic structure of transition metal ion complexes. For $d^{1,9}$ systems such as $Cu^{2+}$ centered complexes with no fine structure the principal values of the ***g***-tensor of the spin Hamiltonian $H = \beta_e B \cdot g \cdot S$, reflecting the symmetry of the ligand field, can be derived from the *EPR* spectrum, where $B$ is the external magnetic field, $S$ is the spin vector, and $\beta_e = g_e \cdot \mu_B$ (*Landé g*-factor for the free electron, $g_e = 2.0023$, *Bohr* magneton $\mu_B$). In this contribution $g$ values for dioptase, $Cu_6Si_6O_{18} \cdot 6H_2O$, determined by *Reddy et al.* [19], and data measured by *Meibohm* [13] for synthetic $Cu_6Ge_6O_{18} \cdot 6H_2O$ were used as expressed in its principal axes system.

## 3. Results and Discussion

Results of a Gaussian deconvolution of the UV-VIS spectra for the hydrated and dehydrated compounds, respectively, are given in Table 1 and depicted in Figures 2a to 2d. $\lambda$(nm) and $\Gamma$(nm) represent wavelength and full width of the excitation peaks, and the wavenumber E(cm$^{-1}$) denotes the excitation energy. The remarkable oscillator strength $f$ (given in arbitrary units) is the consequence of non-zero *dd* transition probabilities due to the absence of symmetry elements on the Cu position with $C_1$ site symmetry and the disphenoidic (stocky tetrahedral) oxygen environment with 4 distinct equatorial bond lengths indicating $Cu_{3d}$-$O_{2p}$ hybridization. The relative width $\Gamma/\lambda$ of the bands of the dehydrated compounds is about 18%, whereas that of the hydrated ones suffer additional broadening to about 23% caused by a vibronic contribution of the water molecule rings and due to assumed peak overlapping according to the below presented assignment.

The steep increase of absorption at the badly resolved high energy limit of the *UV-VIS* spectra has been simulated by a Gaussian curve, too, and may be interpreted as absorption edge, the large gap between valence and conduction band of isolator compounds. The gap is determined around 3.80 eV for dioptase and shifts to 3.76 eV for Ge-dioptase, respectively. It is slightly lower for the dehydrated compounds, giving 3.52 and 3.47 eV, respectively (Table 1). For comparison, *Rudko* [24] observed an absorption edge near 3.5 eV for the charge transfer insulator $CuGeO_3$. The absorption structures at high energy just before the energy gap may be attributed to simultaneous ligand field transitions involving both metal centres of the dimer at twice the monomer transition energy (*SPE*), because their oscillator strengths are too weak for charge transfer (*CT*) transitions.

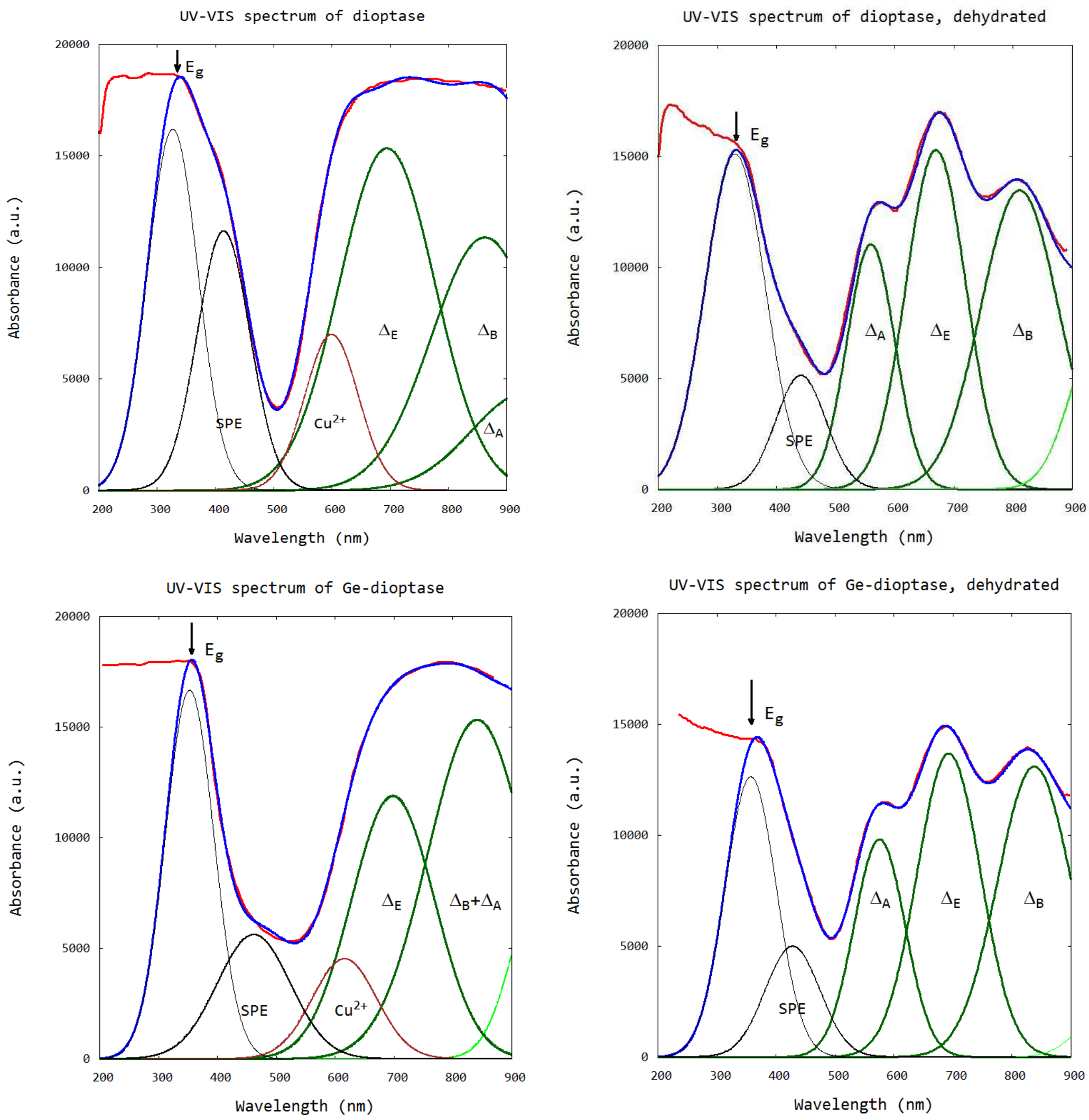

**Figure 3**. UV-VIS spectra of a) dioptase, b) dioptase dehydrated, c) Ge-dioptase, d) Ge-diptase dehydrated. Measured spectra red, calculated spectra blue.

The color of $Cu^{2+}$ compounds with their *Jahn-Teller* distorted coordination polyhedra [25] is the conspicuously recognized property of this transition metal ion and is attributed to electronic excitations between its d-orbitals. The coordination polyhedron of copper in the $d^9$ state with the unpaired electron in the $x^2$-$y^2$ orbital is an elongated octahedron leading to splitting of formerly degenerated d-states. A recently found impressive example for a *Jahn-Teller* elongated octahedron is the new prototypic crystal structure of tetragonal CuO with a $c > a$ rock salt structure [26].

The transition energies $\Delta_n$ ($cm^{-1}$), derived from broad Gaussian-shaped absorption bands of *UV-VIS* spectra, are the energy differences between the $^2B_{1g}(x^2\text{-}y^2)$ ground state and the $^2B_{2g}(xy)$, $^2A_{1g}(z^2)$ and $^2E_g(xz,yz)$ excited states and can be connected with crystal field splitting parameters representing orbital energies. Bearing in mind the $Cu^{2+}$ site symmetry of $D_{4h}$ or lower, we are faced with an equatorial $Dq_e$ splitting parameter and two radial *D*s and *D*t ones (*Gerloch and Slade*, [27]). The crystal field theory (*CFT*) allows for the following relations:

$$\Delta_B = {}^2B_{2g}(xy) \; - \; {}^2B_{1g}(x^2\text{-}y^2) \; = 10Dq_e \qquad (1)$$

$$\Delta_E = {}^2E_g(xz,yz) - {}^2B_{1g}(x^2\text{-}y^2) = 3Ds + 10Dq_e - 5Dt. \quad (2)$$
$$\Delta_A = {}^2A_{1g}(z^2) - {}^2B_{1g}(x^2\text{-}y^2) = 4Ds + 5Dt \quad (3)$$

Conversely, the $D$ parameters can be recalculated as

$$Dq_e = \Delta_B/10 \quad (4)$$
$$Ds = (\Delta_A + \Delta_E - \Delta_B)/7 \quad (5)$$
$$Dt = (3\Delta_A + 4(\Delta_B - \Delta_E))/35 \quad (6)$$

Whereas $\Delta_E$ is always moderately larger than $\Delta_B$, $\Delta_A$ ranges from about 8500 $cm^{-1}$($< \Delta_B$) for shortest axial bonds to at least 21500 $cm^{-1}$ ($> \Delta_E$) for axially non-existent bonds (squared-planar coordination). A more quantitative description of ligand field parameters using effective charges and bond lengths results in the following relations [26]:

$$Dq_e = \eta_e \cdot Z_L \cdot e^2\langle r^4\rangle/(6 \cdot R_e^5), \quad Dq_a = \eta_a \cdot Z_L \cdot e^2\langle r^4\rangle/(6 \cdot R_a^5) \quad (7)a,b$$
$$Cp_e = 2\eta_s \cdot Z_L \cdot e^2\langle r^2\rangle/(7 \cdot R_e^3), \quad Cp_a = 2\eta_s \cdot Z_L\, e^2\langle r^2\rangle/(7 \cdot R_a^3) \quad (8)a,b$$
$$Ds = Cp_e - Cp_a \quad (9)$$
$$Dt = 4/7 \cdot \eta_t\,(Dq_e - \eta_l \cdot Dq_a) \quad (10)$$

where $\eta \cdot Z_L$ represents an effective ligand charge, $R_e$ and $R_a$ are equatorial and axial bond lengths in Å, and $\langle r^4\rangle = 0.214$ Å$^4$ is the mean value of the fourth power of a 3*d* orbital radial distance from the nucleus, respectively $\langle r^2\rangle = 0.294$ Å$^2$ the mean of the second power of the radial distance. For $\langle r^n\rangle$ the values calculated by *Haverkort* within the *Hartree-Fock* approximation are used [28]. Because $\langle r^4\rangle$ is a measure proportional to the Cu(II) effective nuclear charge, one should multiply this value by a factor of 4 to give a realistic value of about 0.85 for the *Scott* charge, which would represent 42.5% ionicity of the Cu-O bond.

**Table 1**. Results of the Gaussian profile deconvolution of the UV-VIS spectra of the dioptase family. $f$ oscillator strength (arbitrary units), $\Gamma$(nm) full band width at half $f$, $E$($cm^{-1}$) band energy, *SPE* simultaneous pair excitation, $E_g$ large energy gap, $Cu^{2+}$ fluorescence contribution.

| $Cu_6Si_6O_{18}\cdot 6H_2O$ (dioptase) | | | | | $Cu_6Ge_6O_{18}\cdot 6H_2O$ (Ge-dioptase) | | | | |
|---|---|---|---|---|---|---|---|---|---|
| $f$ | $\lambda$ (nm) | $\Gamma$(nm) | $E$($cm^{-1}$) | Assignment | $f$ | $\lambda$ (nm) | $\Gamma$(nm) | $E$($cm^{-1}$) | Assignment |
| 329 | 935 | 208 | 10700 | $\Delta_A$ | 1000 | 842 | 198 | 11884 | $\Delta_B + \Delta_A$ |
| 815 | 869 | 208 | 11507 | $\Delta_B$ | | | | | |
| 1000 | 695 | 194 | 14400 | $\Delta_E$ | 645 | 698 | 166 | 14321 | $\Delta_E$ |
| 256 | 600 | 109 | 16670 | $Cu^{2+}$ | 185 | 616 | 129 | 16230 | $Cu^{2+}$ |
| 421 | 414 | 108 | 24160 | *SPE* ? | 271 | 462 | 147 | 21650 | *SPE* ? |
| (554) | 326 | 102 | 30660 | $E_g$ | (531) | 353 | 97 | 28350 | $E_g$ |
| $Cu_6Si_6O_{18}$ (dioptase dehydrated) | | | | | $Cu_6Ge_6O_{18}$ (Ge-dioptase dehydrated) | | | | |
| $f$ | $\lambda$ (nm) | $\Gamma$(nm) | $E$($cm^{-1}$) | Assignment | $f$ | $\lambda$ (nm) | $\Gamma$(nm) | $E$($cm^{-1}$) | Assignment |
| 1000 | 811 | 158 | 12330 | $\Delta_B$ | 1000 | 827 | 155 | 12100 | $\Delta_B$ |
| 866 | 668 | 120 | 14960 | $\Delta_E$ | 843 | 679 | 127 | 14723 | $\Delta_E$ |
| 492 | 558 | 96 | 17930 | $\Delta_A$ | 450 | 565 | 97 | 17700 | $\Delta_A$ |
| 235 | 441 | 100 | 22680 | *SPE* ? | 355 | 437 | 108 | 22880 | *SPE* ? |
| (869) | 330 | 120 | 30300 | $E_g$ | (630) | 357 | 102 | 28000 | $E_g$ |

Quoting *Gerloch and Slade* [27] ones more, in the crystal-field theory with its point-charge formalism charges as well as bond lengths have to be considered as effective parameters that are not independent of each other. Therefore, cationic and ligand charges should be combined to common adaptable factors $Q_e^2 = \eta_e \cdot Z_L \langle r^4 \rangle$ respectively $Q_s^2 = \eta_s \cdot Z_L \langle r^2 \rangle$.

For comparison of calculated band energies with experimental ones given in $cm^{-1}$ an energy conversion factor $c_E = e^2/(4\pi\varepsilon_o \cdot 1(\text{Å})) = 1.1608 \cdot 10^5$ is applied.

*Lebernegg et al.* [29] found no general theoretical justification for $R^{-5}$ dependence of ligand-field splitting. Nevertheless, one can use the inverse fifth power relationship $Dq_e \propto R^{-5}$ in order to calculate a linear regression curve of $Dq_e$ (or $\Delta_B$) values against the mean of the four equatorial copper-oxygen distances $R_e$(Å) according to Equation 7a for selected compounds with a reduced connectedness with respect to equatorial sharing, at the beginning excluding sheet structures as exemplified by cuprates.

The plot is depicted in Figure 3 and extrapolates well through the origin with $Q_e^2 = 1.723$, giving effective charge numbers of $Q_e = \pm 1.313$ assumed evenly distributed over $Cu^{2+}$ and ligands. The calculated $Dq_e$ values deviate less than 1.6% from the experimental ones.

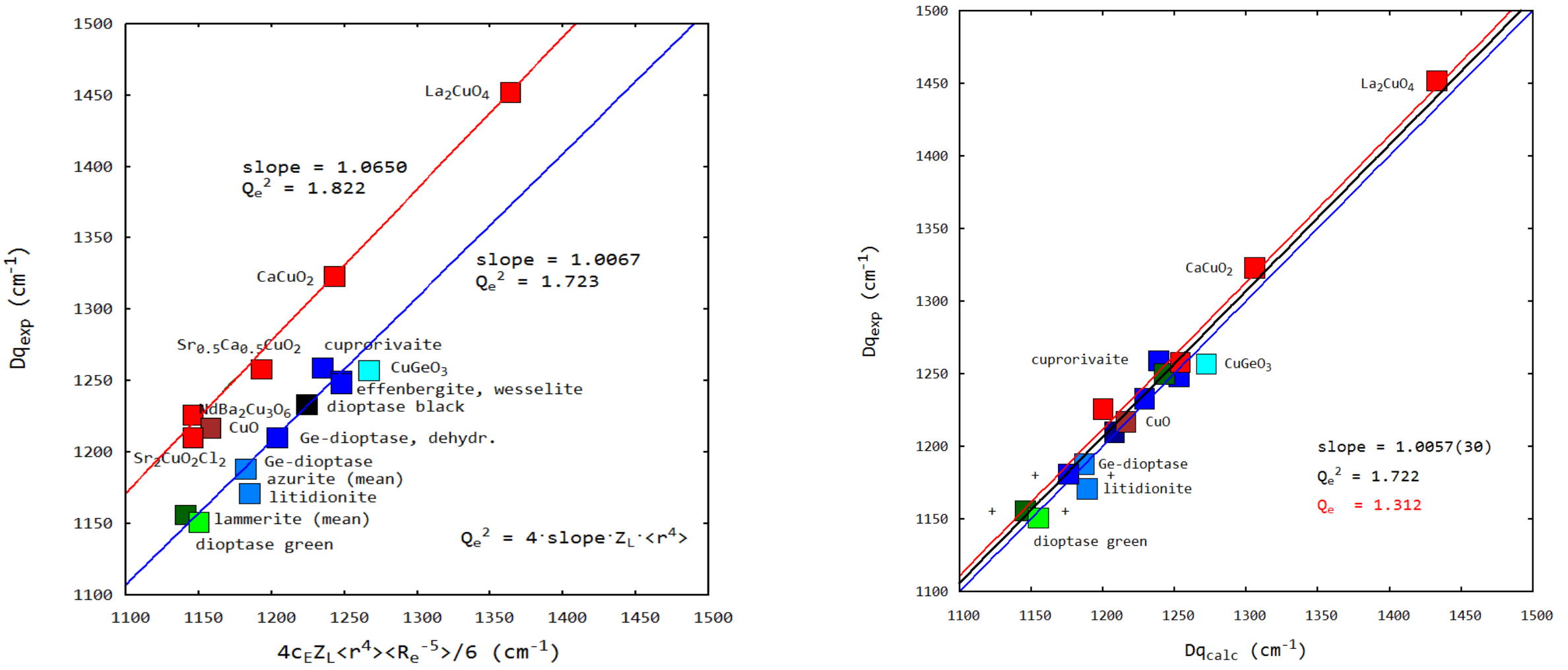

**Figure 4**. Calculated *Dq* energies versus experimental ones for dioptase and related compounds. In the right plot, an additional magnetic contribution of $Dq_{calc}$ was considered for cuprates.

We chose compounds of the *Egyptian Blue* family (cuprorivaite, wesselite, effenbergite,) with isolated $D_{4h}$ plaquettes, the dehydrated dioptase compounds with equatorially edge-shared dimers, further connected via water oxygen to corner-shared spiral chains in the fully hydrated compounds, litidionite as characterized by pyramid-edge-shared dimers (cis-arrangement), in contrast to lammerite with infinite chains of such units and with two distinct Cu sites, further azurite with ‘octahedral’ chains (two distinct sites), and finally conichalcite and $CuGeO_3$ showing infinite single chains with equatorially edge-shared ‘octahedra’. One may learn more about the structural hierarchy of special copper oxy-salt minerals from *Eby and Hawthorne* [30].

Recently, the energy and symmetry of dd excitations of some undoped layered cuprates have been measured by $CuL_3$ resonant *X*-ray scattering [31]. The well assigned *dd* excitations of these compounds with high connectedness were found to be higher than the energies of the compounds described before. Two further compounds can be added to this group with due allowance, the ‘green’ ino-cuprate phase $Y_2BaCuO_5$ and multiferroic CuO as limiting case. Applying Eq. 7a, a steeper slope with $Q_e^2 = 1.82$ results, representing higher excitation energies and effective charges ($Q_e \pm 1.35$) than for the dioptase group. Too low a bond strength of the ‘green’ phase $Y_2BaCuO_5$ is striking (Table 2) and a distinct $Q_e^2$ has been applied (Table 4).

The different connectedness of cuprates in comparison to the dioptase group is manifested in a larger contribution of the principal magnetic super-exchange interaction $J_z$ to the optical excitation energies. In Figure 5, this contribution is depicted versus the Cu–O–Cu bond angle $\Phi$, a representation first used by *Rocquefelte et al.* [32], and here applied in an extended form, illustrating both dioptase group compounds and cuprate ones. A data fit resulted in the relation

$$J_z(\Phi) = 0.091\cdot(\Phi\text{-}90)^{1.652}\ (\text{meV}) = 0.734\cdot(\Phi\text{-}90)^{1.652}(\text{cm}^{-1}), \qquad 11$$

with an exponent near 5/3, explained by chemical pressure (*Rocquefelte et al.*, 2012) [32][33]. Adding $J_z(\Phi)\cdot\ln(2)$ as bond angle dependent contribution to the bond length dependent one, a surprisingly good agreement is achieved between the two groups of compounds, now giving $Q_e^2$ = 1.722, respectively $Q_e$ = 1.312. It should be noticed that for the dioptase group an antiferromagnetic contribution is not included, because $T_N$ is lower than room temperature, at which the optical spectra are taken.

**Table 2.** Coordination numbers *CN*, bond length and bond valence sums **s** for selected Cu(II) compounds $s_e$ equatorial sum, $s_a$ axial sum, $\Sigma s$ overall sum (particularly striking values in red).

| Compound | *CN* | *d*(Cu-O) (Å) | Bond strength $s_e$ | Bond strength $s_a$ | Bond strength $\Sigma s$ | Reference |
|---|---|---|---|---|---|---|
| $Ca_{0.5}Sr_{0.5}CuO_4$ | 4 | 1.945 | 1.913 | - | 1.913 | [39] |
| $CaCuO_2$ | | 1.928 | 2.004 | - | 2.004 | [33] |
| $BaCuSi_4O_{10}$ Effenbergite | | 1.925 | 2.022 | - | 2.022 | [40] |
| $SrCuSi_4O_{10}$ Wesselite | | 1.925 | 2.022 | - | 2.022 | [41] |
| $CaCuSi_4O_{10}$ Cuprorivaite | | 1.929 | 1.998 | - | 1.998 | [42] |
| $Cu_6Si_6O_{18}$ | 4 + (1) | 1.9250 1.9294 1.9354 1.9466 3.3153 | 1.969 | 0.023 | 1.992 | [12] |
| $Cu_6Ge_6O_{18}$ | | 1.9043 1.9284 1.9380 1.9979 3.3841 | 1.954 | - | 1.954 | [16] |
| $Y_2BaCuO_5$ | 2 + 2 + 1 | 1.985 1.988 2.206 | 1.692 | 0.232 | 1.923 | [43] |
| $CuGeO_3$ | 4 + 2 | 1.941 2.926 | 1.941 | 0.093 | 2.022 | [1] [2] [3] |
| $Cu_6Si_6O_{18}\cdot 6H_2O$ | 6 | 1.952 1.952 1.959 1.983 2.502 2.648 | 1.818 | 0.195 | 2.014 | [44] |
| $Cu_6Ge_{5.4}Si_{0.6}O_{18}\cdot 6H_2O$ | 6 | | | | | [16] |
| $Cu_6Ge_6O_{18}\cdot 6H_2O$ | 6 | 1.9037 1.9486 1.9547 1.9884 2.6364 2.6696 | 1.894 | 0.159 | 2.053 | [10] |
| $Cu_3(CO_3)_2(OH)_2$ Azurite | 2 + 2 + 2 Cu(1) site | 1.9387 1.9455 2.9840 | 1.953 | 0.083 | 2.036 | [45] |
| | 6 Cu(2) site | 1.9385 1.9388 1.9675 1.9947 2.3608 2.7578 | 1.830 | 0.223 | 2.053 | |
| $CuSO_4\cdot 5H_2O$ Chalcantite | 2+2+2 Cu(1) site | 1.9748 1.9770 2.3858 | 1.769 | 0.250 | 2.019 | [46] |
| | 2+2+2 Cu(2) site | 1.9447 1.9696 2.4400 | 1.739 | 0.293 | 2.033 | |
| $KNaCuSi_4O_{10}$ Litidionite | 6 | 1.9220 1.9434 1.9683 1.9799 2.6238 3.4024 | 1.863 | 0.109 | 1.972 | [21]; this work |
| $Cu_3(AsO_4)_2$ Lammerite | 2 + 2 + 2 Cu(1) site | 1.933 1.974 2.923 | 1.864 | 0.093 | 1.957 | [47] |
| | 6 Cu(2) site | 1.941 1.947 1.972 2.028 2.282 2.782 | 1.772 | 0.254 | 2.026 | |
| $CaCuAsO_4OH$ Conichalcite | 6 | 1.8850 1.8855 2.0666 2.0688 2.2976 2.3882 | 1.811 | 0.332 | 2.143 | [48]; this work |
| $La_2CuO_4$ | 4 + 2 | 1.9043 2.4145 | 2.151 | 0.278 | 2.439 | [49] |
| $Sr_2CuO_2Cl_2$ | 4 + 2 | 1.9864 2.860 | 1.687 | 0.292 | 1.979 | [50] |

**Table 3**. Collection of some properties and data for selected Cu(II) oxo-compounds

$\Delta_B$ = $^2B_{2g}$(xy) - $^2B_{1g}$($x^2$-$y^2$), $\Delta_E$ = $^2E_g$(xz,yz) - $^2B_{1g}$($x^2$-$y^2$), $\Delta_A$ = $^2A_{1g}$($z^2$) - $^2B_{1g}$($x^2$-$y^2$), energies in $cm^{-1}$. Other transitions: *CT* charge transfer, *SPE* simultaneous pair excitation, $E_g$ energy gap, $Cu^{2+}$ fluorescence peak.

<table>
<tr><th>Compound</th><th>Color</th><th>Cu site symmetry</th><th>$\Delta_B$</th><th>$\Delta_E$</th><th>$\Delta_A$</th><th>Other Transition</th><th>Reference</th></tr>
<tr><td>$CaCuO_2$<br>(infinite layer)</td><td></td><td>$D_{4h}$</td><td>13230</td><td>15730</td><td>21370</td><td>-</td><td>[31]</td></tr>
<tr><td>Tenorite<br>CuO</td><td>brownish</td><td>$C_i$</td><td>12170</td><td>12930<br>15530</td><td>16670</td><td>many</td><td>[51] **</td></tr>
<tr><td>Conichalcite synth.<br>$CaCuAsO_4OH$</td><td>light green</td><td>$C_1$</td><td>10575</td><td>12500</td><td>8585</td><td>15313 $Cu^{2+}$<br>31370 (**$E_g$**)</td><td>[52]</td></tr>
<tr><td>Lammerite synth.<br>$Cu_3(AsO_4)_2$</td><td>dark green</td><td>$C_1$<br>$C_i$</td><td>11530</td><td>12400<br>14050</td><td>10200<br>14050</td><td>23260 CuO<br>31250 (**$E_g$**)</td><td>[22]; this work *</td></tr>
<tr><td rowspan="2">$Y_2BaCuO_5$</td><td rowspan="2">vivid green</td><td rowspan="2">$D_{4h}$</td><td colspan="2">12500</td><td rowspan="2">14700</td><td rowspan="2">25970 CT</td><td rowspan="2">[53] **</td></tr>
<tr><td>11700</td><td>13200</td></tr>
<tr><td>Azurite<br>$Cu_3(CO_3)_2(OH)_2$</td><td>blue</td><td>$C_i$<br>$C_1$</td><td>11806<br>11806</td><td>16484<br>11806</td><td>16484<br>11806</td><td>19793<br>17952</td><td>[54] **<br>*T* = 80 K</td></tr>
<tr><td>Chalcantite<br>$CuSO_4 \cdot 5H_2O$</td><td>deep blue</td><td>$C_1$<br>$C_1$</td><td>11407<br>11860</td><td>13308<br>13488</td><td>9699<br>9735</td><td>15234 $Cu^{2+}$<br>15567 $Cu^{2+}$</td><td>[55]</td></tr>
<tr><td>Litidionite synth.<br>$KNaCuSi_4O_{10}$</td><td>light blue</td><td>$C_i$</td><td>11723</td><td>14700</td><td>13900</td><td>21400 *SPE*<br>31600 (**$E_g$**)</td><td>[22] [23]; this work</td></tr>
<tr><td>$Cu_6Ge_6O_{18}$</td><td>dirty blue</td><td>$C_1$</td><td>12100</td><td>14723</td><td>17700</td><td>22880 *CT*</td><td>[16];<br>this work</td></tr>
<tr><td>$Cu_6Ge_6O_{18} \cdot 6H_2O$</td><td>bluish-green</td><td>$C_1$</td><td>11880</td><td>14320</td><td>11880</td><td>16230 $Cu^{2+}$<br>21650 *CT*<br>30300 (**$E_g$**)</td><td>[16];<br>this work</td></tr>
<tr><td>$Cu_6Ge_{5.4}Si_{0.6}O_{18}$<br>$\cdot 6H_2O$</td><td>dark green</td><td>$C_1$</td><td colspan="2">13300</td><td>19400</td><td>23900 *CT*</td><td>[17]</td></tr>
<tr><td>$CuGeO_3$</td><td>turquoise</td><td>$D_{2h}$</td><td>12570</td><td>13970</td><td>12920</td><td>15733 $Cu^{2+}$<br>15800 *CT*</td><td>[24]; this work</td></tr>
<tr><td>$BaCuSi_4O_{10}$ synth.<br>(Effenbergite)</td><td rowspan="3">deep blue</td><td rowspan="3">$D_{4h}$</td><td>12200</td><td>15950</td><td>18520</td><td>-</td><td>[56]</td></tr>
<tr><td>$SrCuSi_4O_{10}$ synth.<br>(Wesselite)</td><td>12480</td><td>16050</td><td>18520</td><td>-</td><td>[56]</td></tr>
<tr><td>$CaCuSi_4O_{10}$ synth.<br>(Cuprorivaite)</td><td>12590<br>12740</td><td>15760<br>16130</td><td>18530<br>18520</td><td>-</td><td>[43]<br>[57]</td></tr>
<tr><td>Dioptase dehydrated<br>Dioptase partly dehydrated</td><td>black<br>dark blue</td><td>$C_1$</td><td>12330<br>12500</td><td>14960<br>14500</td><td>17930<br>17600</td><td>22680 *CT*<br>-</td><td>This work;<br>[19]</td></tr>
<tr><td rowspan="2">Dioptase<br>$Cu_6Si_6O_{18} \cdot 6H_2O$</td><td rowspan="2">emerald green</td><td rowspan="2">$C_1$</td><td>11500<br>12495</td><td>14500<br>15010</td><td>17000<br>10200</td><td></td><td>[18]<br>[19]</td></tr>
<tr><td>11507</td><td>14400</td><td>10700</td><td>16670 $Cu^{2+}$<br>24160 *CT*<br>30660 (**$E_g$**)</td><td>This work</td></tr>
</table>

* synthetic lammerite with an amount of CuO

** for this work a different assignment as given in the reference was used

The $R^{-5}$ inverse power of Cu-O bond lengths is nearly a measure for the bond strength. Therefore, the reliability of the fit can be enhanced applying the empirical Cu-O bond strength relation $s = \Sigma(R/R_0)^{-N}$ [34] by choosing only the bond strength sum $s_e$ of the four equatorial bonds. New values $R_0 = 1.713(9)$ Å, $N = 5.76(16)$ were re-calculated for this work [35]. Results of a double-regression yielded for the cuprate group

$$\Delta_B\ (\mathrm{cm}^{-1}) = (6658 \pm 38)\cdot s_e \quad \text{and} \quad \Delta_B\ (\mathrm{cm}^{-1}) = (6269 \pm 21)\cdot (s_e + 1.38\cdot 10^{-4} J_z) \qquad 12$$

for both the dioptase group and cuprates, the last mentioned corrected by a bond angle dependent (magnetic) contribution (Figure 6). It is recommended to extend the analytic bond strength – bond length expression by a magnetic (angle dependent) contribution. In contrast to this result, the quoted authors [31] fitted their cuprate data with a lower slope of $N = 4.2$. On the other hand, the selection of compounds for such fit is not convincing, because an influence of some equatorial $O^{1-}$ ions in $La_2CuO_4$ (high bond strength, see Table 2) on the excitation energies can be expected. In addition, the epitaxially grown infinite-layer structure of $Ca_{0.5}Sr_{0.5}CuO_2$ is obviously strained.

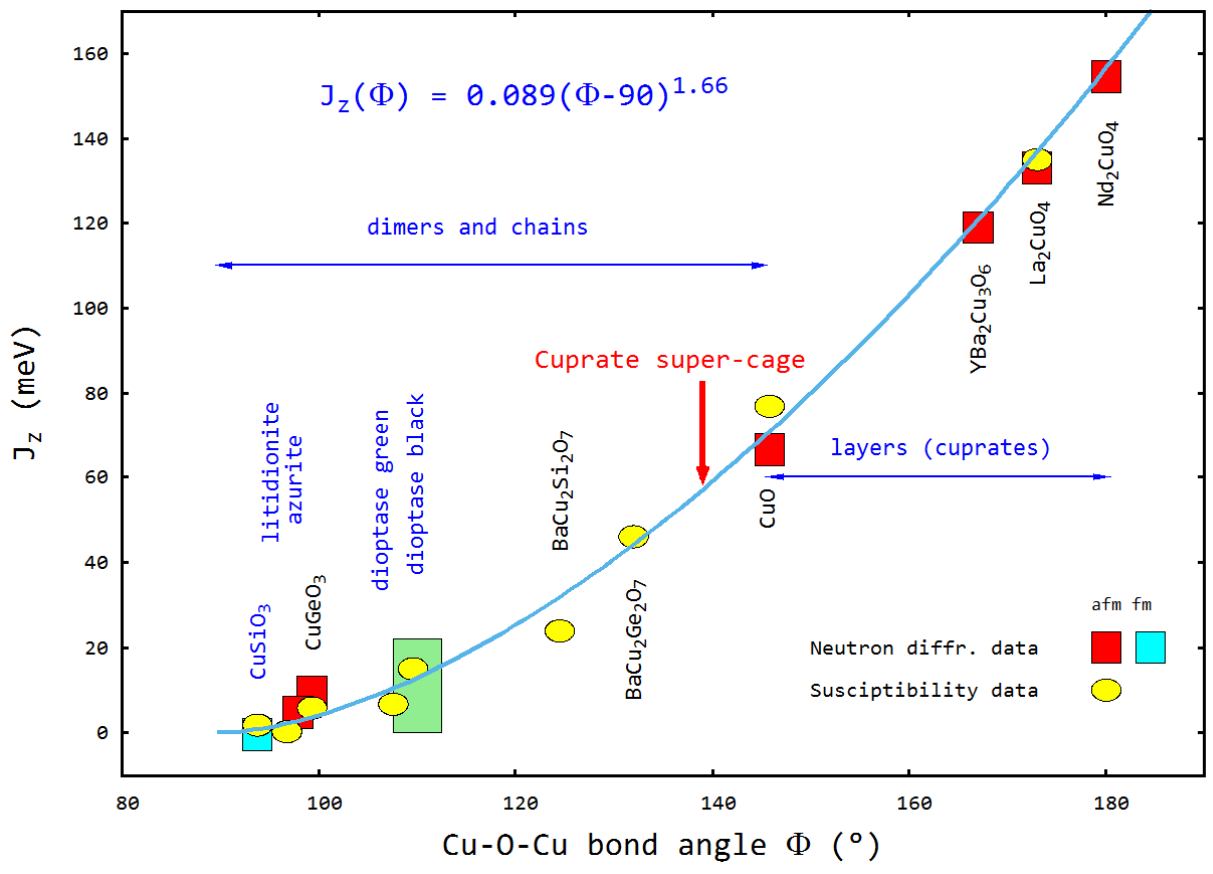

**Figure 5**. Principal superexchange interaction $J_z$ versus Cu–O–Cu bond angle $\Phi$ [32,33].

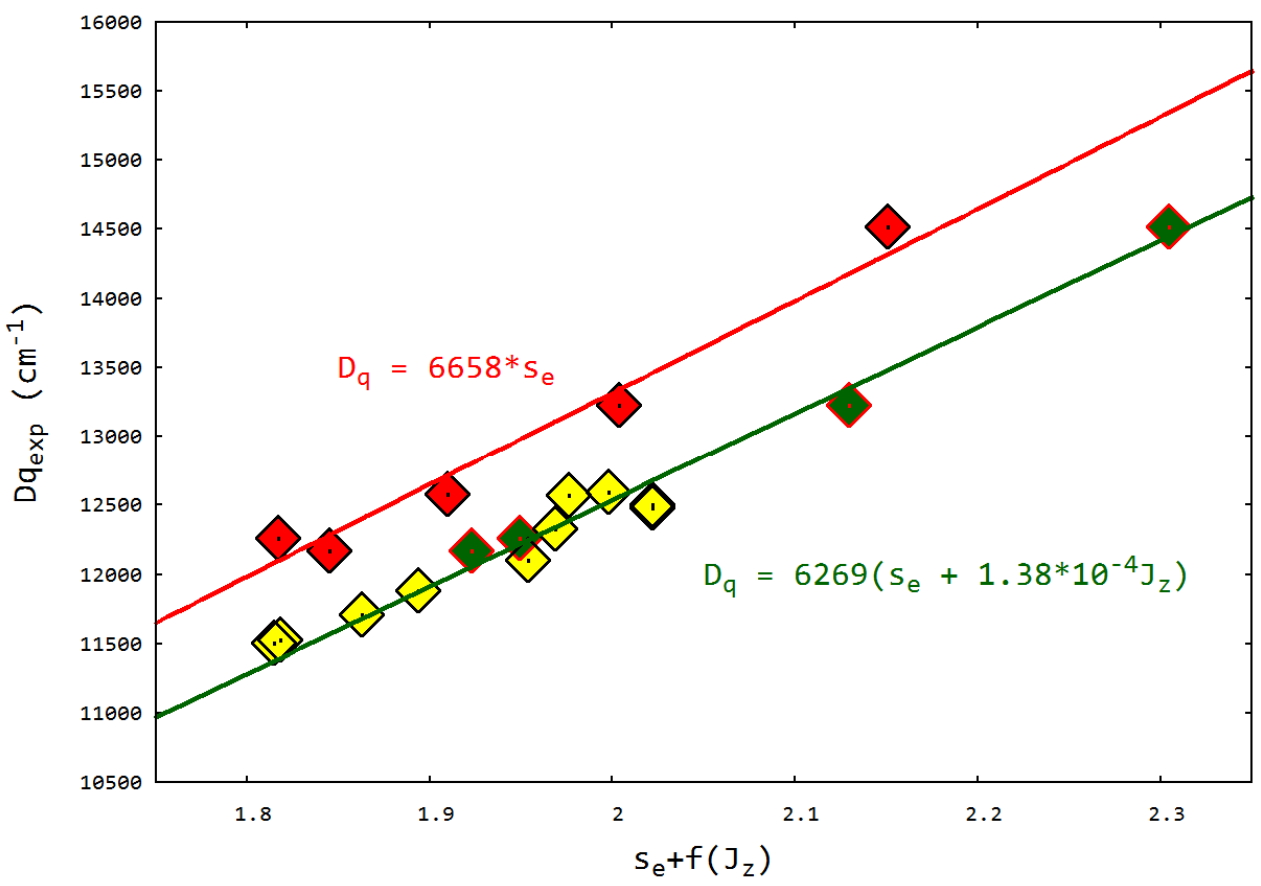

**Figure 6.** Linear relation between $D$q (cm$^{-1}$) and the equatorial bond strength $s_e$, in case of cuprates (red curve) corrected by a bond angle dependent (magnetic) contribution to show a single linear plot (green curve) with dioptase group compounds (yellow) besides cuprates (now green).

Turning now to the calculation of splitting parameters $D$s (Eq.9) and $D$t (Eq.10) involving axial ligands one has to distinguish between pure oxo-ligands and more ionic ones like $H_2O$ and $Cl^{1-}$ with increased cationic charge and assumed higher $Dq_a$ values. $H_2O$ (as equatorial ligands) are found in chalcantite, and hydroxyl groups in azurite and conichalcite; the last compound has the most distorted 'octahedron' and should show a pronounced splitting of the $^2E_g$ term, which is not considered here.

In the case of square-planar environment it is useful to limit the extent of the $d_{z^2}$ orbital with 'long' auxiliary axial bonds. Dehydrated dioptase and the germanate analogue already have some far distant oxygen ions (see Table 2) within the $d_{z^2}$ orbital sphere of influence. For the group of $M^{2+}CuO_2$ layered cuprates the limit is given by the layer separation down *c* of about 3.3 Å. Again the results differ somewhat for the two groups of compounds with slightly different effective charges. From the fitted values for *D*s and *D*t the $\Delta_E$ (Equation 2) and $\Delta_A$ (Equation 3) energies have been calculated as well as $^2E_g$ and $^2A_{1g}$. Results are summarized in Table 4.

In order to check the correct assignment one can use a relation between experimental $B_{2g}$, $E_g$ and $A_{1g}$ values of the form $f(\Delta) = \Delta_A/(2\cdot\Delta_E\text{-}\Delta_B) = 2/3$, which results from equations 1 to 3. Obviously this relation holds only for shortest axial bonds and more octahedral ligand environment, such fulfilling the precondition for the underlying ionic model. For non-existent axial bonds the value for the quotient is close to unity. The values listed in Table 4 indicate clearly a bond length dependence. An empirical function $f(R) = \alpha_1\cdot(R_a/R_e)$ may serve as a correction giving quotients $f(\Delta)/f(R)$ near unity when using $\alpha_1 = 0.59$ for dioptase group compounds respectively 0.70 for cuprates. Compounds with axial water ligands or $Cl^{1-}$ can clearly be identified by relatively small values. Another possibility here published the first time ever is to use the linear relation

$$\Delta_A = \alpha\cdot\left(\frac{<R_a>}{<R_e>} - \varepsilon\right) \quad \text{, where } \alpha = (17892 \pm 60)\ \text{cm}^{-1}, \qquad 13$$

with a $R_a/R_e$ ratio including well adapted auxiliary $R_a$ bonds for compounds of coordination number 4, but different $\varepsilon$ values for the dioptase group ($\varepsilon = 1/\sqrt{2}$) and cuprates ($\varepsilon = 1/2$) to guide the regression line well through the origin (Figure 7 and Table 5).

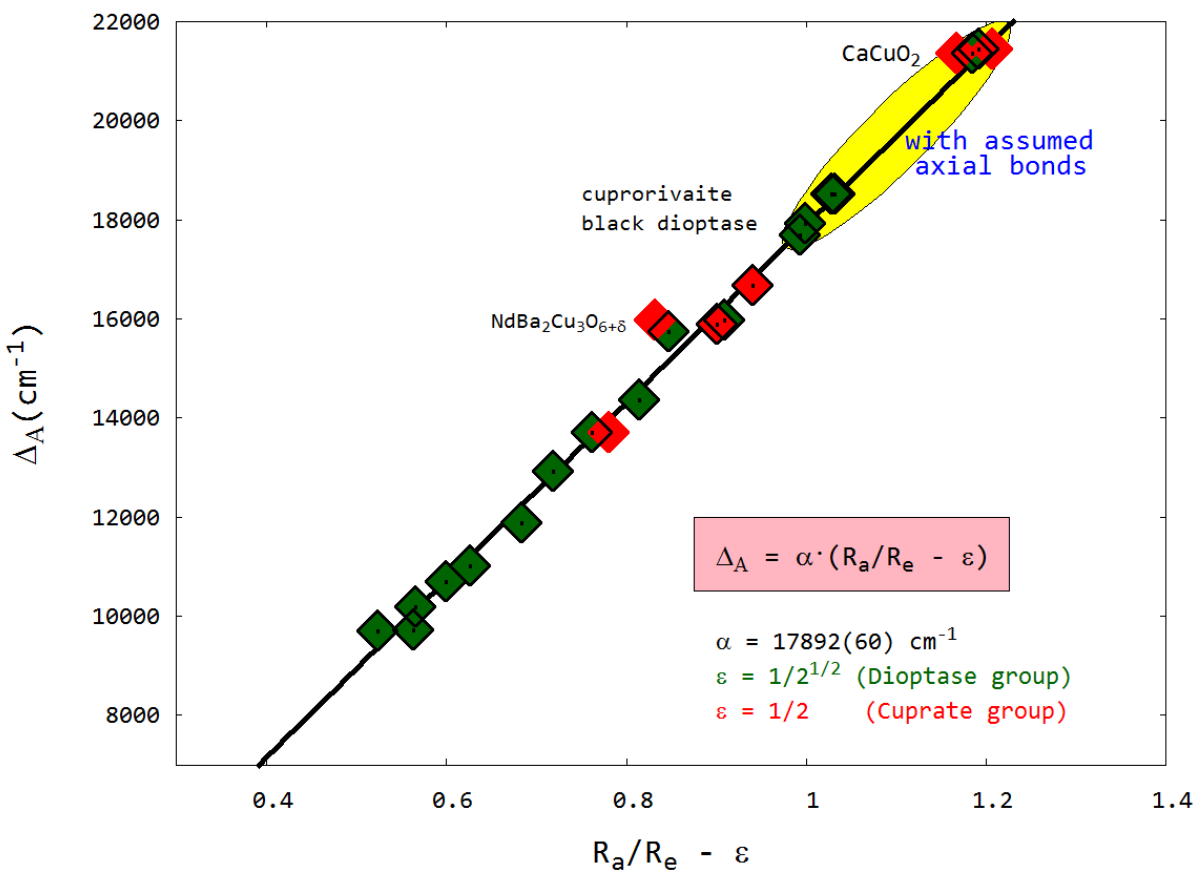

**Figure 7**. $\Delta_A$ excitation energies ($cm^{-1}$) depicted versus a function of axial to equatorial bond distances. Again the cuprate group excitations (in red) must be corrected by a (magnetic) contribution to reliably represent all data in a single regression line. Auxiliary axial bonds (see yellow field) were introduced in case of compounds with really missing axial bonds (coordination number 4).

It should be stressed with respect to the use of mean bond distances in equation 13 that also in the equations 7 to 10 the mean of corresponding bond distances is taken first and then their inverse fifth power is calculated to yield the convincing results of Table 4.

**Table 4**. Comparison of experimental and calculated excitation energies and orbital ones in $cm^{-1}$.

$\Delta_B$ is sorted from high values to low ones down the table; $f(\Delta) = \Delta_A/(2\cdot\Delta_E - \Delta_B)$, $f(R) = \alpha_1 \cdot (R_a/R_e)$ . Experimental and calculated $\Delta$ values are arranged one above the other.

**Isolated CuO plaquettes, clusters and chains:** Calculation with $Q_e^2 = 1.733$, $Q_s^2 = 0.761$, $\eta_s = 1.276$, $\eta_t = 1.738$, $\alpha_1 = 0.59$

| Phase | $\Delta_B$ | $\Delta_E$ | $\Delta_A$ | *Ds* | *Dq* | *Dt* | *Dt*/*Dq* | *Dt*/*Ds* | *f*(Δ) | *f*(Δ)/*f*(*R*) | Ligands equat. | Ligands axial |
|---|---|---|---|---|---|---|---|---|---|---|---|---|
| Cuprorivaite | 12590<br>12501 | 15760<br>15446 | 18530<br>18414 | 3100<br>3051 | 1259<br>1250 | 1226<br>1241 | 0.974 | 0.395 | 0.979 | 0.956 | oxygen | no |
| $CuGeO_3$ | 12570<br>12439 | 13970<br>14209 | 13970<br>14324 | 2196<br>2299 | 1257<br>1244 | 1037<br>1026 | 0.825 | 0.472 | 0.909 | 1.081 | oxygen | oxygen |
| Effenbergite | 12500<br>12566 | 15950<br>15895 | 18520<br>18591 | 3139<br>3132 | 1250<br>1257 | 1139<br>1213 | 0.955 | 0.380 | 0.955 | 0.930 | oxygen | no |
| Wesselite | 12480<br>12566 | 16050<br>15895 | 18520<br>18591 | 3156<br>3132 | 1248<br>1257 | 1179<br>1213 | 0.945 | 0.374 | 0.944 | 0.920 | oxygen | no |
| Dioptase black | 12330<br>12324 | 14960<br>15057 | 17930<br>17850 | 2937<br>2940 | 1233<br>1232 | 1236<br>1218 | 1.003 | 0.421 | 1.019 | 1.013 | oxygen | no |
| $Cu_6Ge_6O_{18}$ | 12100<br>12123 | 14723<br>14845 | 17700<br>17606 | 2903<br>2904 | 1210<br>1212 | 1217<br>1198 | 1.006 | 0.419 | 1.020 | 1.017 | oxygen | no |
| Ge-Dioptase | 11884<br>11929 | 14321<br>14434 | 11884<br>12084 | 2046<br>2084 | 1188<br>1193 | 740<br>750 | 0.623 | 0.362 | 0.709 | 0.866 | oxygen | $H_2O$ |
| Azurite (1) | 11806<br>12070 | 16484<br>14162 | 16484<br>15014 | -<br>2444 | 1181<br>1207 | -<br>1048 | 0.770 | 0.317 | 0.809 | 1.024 | $OH^-$ / oxygen | oxygen |
| Azurite (2) | 11550<br>11589 | 12770<br>12924 | 11300<br>11669 | 1789<br>1858 | 1155<br>1159 | 829<br>848 | 0.718 | 0.464 | 0.808 | 1.067 | $OH^-$ / oxygen | oxygen |
| Litidionite | 11723<br>11794 | 14700<br>14465 | 13900<br>13917 | 2411<br>2397 | 1172<br>1179 | 851<br>866 | 0.726 | 0.353 | 0.786 | 0.915 | oxygen | oxygen |
| Lammerite (1) | 11780<br>11735 | 13744<br>13744 | 14356<br>14357 | 2331<br>2328 | 1178<br>1174 | 1006<br>1000 | 0.854 | 0.432 | 0.914 | 0.914 | oxygen | oxygen |
| Lammerite (2) | 11280<br>11231 | 12400<br>12425 | 10200<br>10279 | 1617<br>1639 | 1128<br>1123 | 746<br>745 | 0.662 | 0.461 | 0.754 | 1.023 | oxygen | oxygen |
| Chalcantite (1) | 11407<br>11135 | 12600<br>12302 | 8900<br>8828 | 1442<br>1428 | 1141<br>1113 | 627<br>623 | 0.549 | 0.435 | 0.645 | 0.906 | $H_2O$ | oxygen |
| Chalcantite (2) | 11860<br>11678 | 13488<br>13369 | 9735<br>98108 | 1623<br>1643 | 1186<br>1168 | 648<br>648 | 0.547 | 0.399 | 0.644 | 0.875 | $H_2O$ | oxygen |
| Dioptase green | 11508<br>11491 | 14398<br>14094 | 10700<br>10325 | 1940<br>1847 | 1151<br>1149 | 588<br>587 | 0.511 | 0.303 | 0.619 | 0.801 | oxygen | $H_2O$ |
| Conichalcite | 11400<br>11119 | 12195<br>11922 | 8585<br>8449 | 1340<br>1339 | 1140<br>1141 | 645<br>658 | 0.566 | 0.481 | 0.661 | 0.940 | $OH^-$ | oxygen |
| $Y_2BaCuO_5$ | 10700<br>10732 | 12500<br>13205 | 14700<br>14571 | 2357<br>2435 | 1070<br>1073 | 1054<br>966 | 0.985 | 0.447 | 1.028 | 1.018 | oxygen | oxygen |

**Cuprates (undoped):** Calculation with $Q_e^2 = 1.755$, $Q_s^2 = 0.936$, $\eta_s = 1.587$, $\eta_t = 2.18$, $\alpha_1 = 0.70$

| Phase | $\Delta_B$ | $\Delta_E$ | $\Delta_A$ | *Ds* | *Dq* | *Dt* | *Dt*/*Dq* | *Dt*/*Ds* | *f*(Δ) | *f*(Δ)/*f*(*R*) | Ligands **equat.** | Ligands **axial** |
|---|---|---|---|---|---|---|---|---|---|---|---|---|
| $La_2CuO_4$ | 14516<br>14308 | 17097<br>16833 | 13710<br>13664 | 2327<br>2313 | 1452<br>1431 | 880<br>883 | 0.606 | 0.378 | 0.697 | 0.788 | oxygen | $O^{1-}$ ? |
| $CaCuO_2$ | 13226<br>13322 | 15726<br>15982 | 21370<br>21671 | 3410<br>3476 | 1323<br>1332 | 1546<br>1554 | 1.169 | 0.453 | 1.173 | 0.994 | oxygen | no |
| $Sr_{0.5}Ca_{0.5}CuO_2$ | 12581<br>12744 | 15565<br>15494 | 21452<br>21026 | 3491<br>3397 | 1258<br>1274 | 1498<br>1488 | 1.190 | 0.429 | 1.157 | 0.976 | oxygen | no |
| $NdBa_2Cu_3O_6$ | 12258<br>12372 | 14113<br>13741 | 15968<br>15940 | 2546<br>2473 | 1226<br>1237 | 1157<br>1210 | 0.944 | 0.454 | 1.000 | 1.072 | oxygen | oxygen |
| CuO (tenorite) | 12170<br>12254 | 14230<br>14520 | 16670<br>16878 | 2676<br>2735 | 1217<br>1225 | 1193<br>1188 | 0.981 | 0.446 | 1.023 | 1.027 | oxygen | oxygen |
| $Sr_2CuO_2Cl_2$ | 12097<br>11838 | 14839<br>14756 | 15887<br>16021 | 2661<br>2706 | 1210<br>1184 | 1048<br>1040 | 0.867 | 0.394 | 0.904 | 0.897 | oxygen | $Cl^-$ |

An additional scaling $\eta_l$ between 1.7 and 2.0 (equation 10) is needed to fit the *Dt* values of compounds with $H_2O$ respectively $Cl^{-1}$ as axial bonds. Also $La_2CuO_4$ needs such correction ($\eta_l$ = 1.71) possibly caused by some $O^{1-}$ expected as axial ligands.

**Table 5.** Experimental $\Delta_A$ energies versus calculated ones using the relation: $\Delta_A = 17892 \cdot (\langle R_a \rangle / \langle R_e \rangle - \varepsilon)$ ($cm^{-1}$), $\varepsilon = 1/\sqrt{2}$ for dioptase group and $\varepsilon = ½$ for cuprates. Auxiliary bonds introduced in case of compounds with coordination number *CN* = [4] are underlined.

| **Phase** | *CN* | $\langle R_e \rangle$(Å) | $\langle R_a \rangle$(Å) | $\Delta_A$ (exp.) | $\Delta_A$ (calc.) |
|---|---|---|---|---|---|
| Cuprorivaite | [4] | 1.9307 | <u>3.35</u> | 18530 | 18393 |
| $CuGeO_3$ | [4+2] | 1.9326 | 2.7549 | 12920 | 12854 |
| Effenbergite | [4] | 1.9265 | <u>3.35</u> | 18520 | 18460 |
| Wesselite | [4] | 1.9265 | <u>3.35</u> | 18520 | 18460 |
| Dioptase black | [4] | 1.9340 | <u>3.30</u> | 17930 | 17874 |
| $Cu_6Ge_6O_{18}$ | [4] | 1.9404 | <u>3.30</u> | 17700 | 17777 |
| Ge-Dioptase | [4+2] | 1.9474 | 2.6622 | 11884 | 11808 |
| Azurite (1) | [4+2] | 1.9434 | 2.9840 | 16488 (?) | 14821 |
| Azurite (2) | [4+1+1] | 1.9593 | 2.5143 | 11806 | 10309 |
| Litidionite | [4+2] | 1.9525 | 2.8434 | 13900 | 13404 |
| Lammerite (1) | [4+2] | 1.9529 | 2.9230 | 14350 | 14128 |
| Lammerite (2) | [4+1+1] | 1.9703 | 2.4609 | 10200 | 9695 |
| Chalcantite (1) | [4+2] | 1.9759 | 2.3858 | 8900 | 8952 |
| Chalcantite (2) | [4+2] | 1.9569 | 2.4400 | 9735 | 9657 |
| Dioptase green | [4+2] | 1.9613 | 2.5688 | 10700 | 10735 |
| Conichalcite | [2+2+2] | 1.9640 | 2.3403 | 8585 | 8668 |
| $Y_2BaCuO_5$ | [4+1] | 1.9899 | 2.196+<u>3.90</u> | 14700 | 14754 |
| $La_2CuO_4$ | [4+2] | 1.9043 | 2.4045 | 13710 | 13646 |
| $CaCuO_2$ | [4] | 1.9281 | <u>3.26</u> | 21370 | 21306 |
| $Sr_{0.5}Ca_{0.5}CuO_2$ | [4] | 1.9440 | <u>3.30</u> | 21452 | 21426 |
| $NdBa_2Cu_3O_{7-\delta}$ | [4+1] | 1.9609 | 2.275+<u>3.25</u> | 15968 | 16260 |
| CuO (tenorite) | [4+2] | 1.9558 | 2.7842 | 16670 | 16524 |
| $Sr_2CuO_2Cl_2$ | [4+2] | 1.9864 | 2.8600 | 15887 | 16841 |

Indeed, the connectedness of copper-ligand units, representing the number of shared copper-oxygen polyhedral, should be important for the *dd* excitation energy. Therefore, besides the equatorial ligand sum that are calculated as fit coordinate we used the bond valence sums to check for inconsistent structural details and signs for mixed valences. Copper polygermanate in the *Pbmm* prototypic structure [1,2] shows too high a sum with $\Sigma s$ = 2.08. There is evidence from *EPR* [36], *X*-ray diffraction [37] and *NQR* measurements [38] that copper is statistically out of center of the $CuO_2$ plaquette, in this way the copper bond strength is reduced towards net charge of 2+. Even large thermal displacement ellipsoids indicate structural features that require a careful evaluation. Bond lengths should be corrected for ‘thermal’ displacement, because not less than their inverse fifth power is used in calculations (see for instance [49]).

## 4. EPR Analysis

Finally, the assignment of the *dd* excitations can be compared with results of *EPR* measurements. For $3d^9$ ions in (nearly) tetragonal ligand symmetry one can apply the following two formulae for the principal components $g_{||}$ and $g_\perp$, if the ground state is $^2B_{1g}$.

$$g_{||} = g_e - \frac{8\lambda}{\Delta_B} - \frac{\lambda^2}{{\Delta_E}^2} - \frac{4\lambda^2}{\Delta_B \Delta_E}, \quad 14$$

$$g_\perp = g_e - \frac{2\lambda}{\Delta_E} - \frac{\lambda^2}{{\Delta_E}^2}, \quad 15$$

where $g_e$ = 2.0023 is the *g*-value for the free electron, and $\lambda$ is the spin-orbital coupling parameter, which yields for the free electron $\lambda_o$ = 829 $cm^{-1}$[58].
The *k* values are the spin orbital reduction factors used to scale the coupling parameters to the free electron value, $k = \lambda/\lambda_o$. This parameter reduction is attributed to covalence effects. Table 5 compares the results for dioptase and Ge-dioptase, respectively. Not surprisingly, the found covalence reduction effect is markedly smaller for the copper germanate than for the copper silicate, in accordance with crystal-chemical experience, confirming higher ionicity of the germanate (Table 6). Unfortunately, *EPR* data for the dehydrated compounds were not available.

**Table 6**. *EPR* analysis of dioptase related compounds

| Notation | Dioptase | Ge-Dioptase |
|---|---|---|
| $\Delta_B$ | 11508 | 11884 |
| $\Delta_E$ | 14395 | 14321 |
| $\Delta_A$ | 10700 | 11884 |
| $g_{\parallel}$ | 2.3601 | 2.3780 |
| $g_{\perp}$ | 2.0511 | 2.0970 |
| $\lambda_{\parallel}$ | -504.08 | -545.53 |
| $k_{\parallel}$ | 0.608 | 0.658 |
| $\lambda_{\perp}$ | -346.91 | -662.63 |
| $k_{\perp}$ | 0.418 | 0.799 |

## 5. Conclusions

As shown, a comparative reappraisal of $Cu^{2+}$ *UV-VIS* spectra benefits from a special consideration of crystal-chemically similar groups of compounds, comparing exemplarily the dioptase group, covering minerals as well as synthetic samples, with cuprates. The assignment of *dd* excitations and their representation each on a single curve is possible by attributing a magnetic (bond angle dependent) contribution to the cuprate group. It is recommended to extend the bond strength – bond length relation by a bond angle dependent (magnetic) contribution. Deviations of the linear representation of orbital excitation energies may be helpful to discriminate results of compounds with peculiar orbital features from those with normal behavior. Fortunately, the first done assignment of well resolved spectra of dehydrated dioptase $Cu_6(Ge,Si)_6O_{18}$ served as input data to deconvolute the badly resolved spectra of as-grown $Cu_6(Ge,Si)_6O_{18}\cdot 6H_2O$ samples. At present, the deconvolution of superposed spectra resulting from different Cu sites of a structure is inadequate. However, a pre-calculation of the expected energy levels can serve as input for fitting the experimental spectra. This has been successfully applied to lammerite. It is recommended to take a series of *UV-VIS* spectra step by step over the entire temperature range from hydrated to fully dehydrated dioptase as a didactic tool to follow the energy levels and their correct assignment, thereby simultaneously controlling the crystal water content by *IR* spectroscopy with a device that offers both analytical possibilities.

## Acknowledgment

The author would like to thank colleague *Prof. Bernd Lehmann* for supporting this work by the donation of wonderful dioptase pieces from *Altyn-Tyube*, Kasakhstan. Also my teacher, the late *Prof. Hugo Strunz,* donated dioptase pieces from the *Tsumeb* mine, Namibia.

**Conflict of Interest:** The author declares no conflict of interest.

## References

1. Ginetti, Y. (1954) Structure cristalline du métagermanate de cuivre. Bulletin Soc. Chim. Belg. **63**, 209-216.
2. Völlenkle, H.; Wittmann, A.; Nowotny, H. Zur Kristallstruktur von $CuGeO_3$. *Monatshefte Chemie* **1967**, *98,* 1352-1357.
3. Hase, M.; Terasaki, I.; Uchinokura, K. Observation of the Spin-Peierls transition in linear $Cu^{2+}$ (spin-1/2) chains in an inorganic compound $CuGeO_3$. *Physical Review Letters* **1993**, *70*, 3651-3654.
4. Boucher, J.P.; Regnault, L.P. The Inorganic Spin-Peierls Compound $CuGeO_3$. *Journal de Physique* I, **1996**, *6*, 1939-1966.
5. Otto, H.H.; Meibohm, M. Crystal structure of copper polysilicate, Cu $[SiO_3]$. *Zeitschrift für Kristallographie* **1999**, *214,* 558-565.
6. Baenitz, M.; Geibel, C.; Dischner, M.; Sparn, G.; Steglich, F.; Otto, H.H.; Meibohm, M.; Gipius, A.A. $CuSiO_3$: a quasi-one-dimensional S = 1/2 antiferromagnetic chain system. *Physical Review B* **2000,** *62*, 12201-12205.
7. Wolfram, H.; Otto, H.H.; Cwik, M.; Braden, M.; André, G.; Bourée, G.F.; Baenitz, M.; Steglich, F. Neutron diffraction study of the nuclear and magnetic structure of the quasi-one-dimensional compound $CuSiO_3$ around $T_N = 8$ K. *Physical Review* B **2004,** *69*, 144115-144127.
8. Gros, C.; Lemmens, P.; Choi, K.Y.; Güntherodt, G.; Baenitz,M.; Otto, H.H. Quantum phase transition in the dioptase magnetic lattice. *Europhysics Letters* **2002,** *60*, 276-280.
9. Janson, O.; Tsirlin, A.A.; Schmitt, M.; Rosner, H. *Arxiv* **2010**:1004.3765vl [cond-mat.str-el]
10. Brandt, H.J.; Otto, H.H. Synthesis and crystal structure of $Cu_6[Ge_6O_{18}]\cdot 6H_2O$, a dioptase-type cyclo-germanate. *Zeitschrift für Kristallographie* **1997**, *212*, 34-40.
11. Hase, M.; Ozawa, K.; Shinya, N. Magnetism of $Cu_6Ge_6O_{18}\cdot xH_2O$ (x = 0 ~ 6), a compound of the one-dimensional Heisenberg *S* = ½ model with competing antiferromagnetic interactions. *Physical Rev*iew B **2003**, *68*, 214421.
12. Law, J. M.; Hoch, C.; Kremer, R.K.; Kang, J.; Lee, C.; Wangbo, M.H.; Otto, H.H. Quantum critical behavior in the dioptase lattice: Magnetic properties of $CuMO_3.yH_2O$ (M = Si, Ge; y = 1,0). *Conference on Highly Frustrated Magnetism* **2010**, Baltimoire, USA.
13. Meibohm, M. Zur Kristallchemie und Kristallphysik von neuen Silikaten und Germanaten des Kupfers mit ketten- und ringförmigen Anionen. *Doctoral Thesis* **1999**, TU Clausthal.
14. Otto, H.H. Über natürliche und synthetische Silicate des Kupfers. *Aufschluss* **2000**, *51*, 47-55.
15. Otto, H.H. Zur Kristallchemie von Verbindungen $Me^{II}[Ge(OH)_6 \mid (SO_4)_2] \cdot 3H_2O$. *Naturwissenschaften* **1968**, *55*, 387.
16. Otto, H.H.; Brandt, H.J.; Meibohm, M. Über die Existenz des Kupferpolysilicats $Cu\{uB_1 1_\infty^1\}[^1SiO_3]$. *Beihefte zu European Journal of Mineralogy* **1996**, *8*, 206.
17. Brandt, H.J. Synthese, Kristallstruktur und Eigenschaften neuer, mit Dioptas verwandter Hexacyclogermanate des Bleis und Kupfers. *Clausthaler Geowissenwchaftliche Dissertationen* **1997**, *H52*, TU Clausthal.
18. Bakhtin, A.I. Optical absorption spectra of $Cu^{2+}$ ions in dioptase. *Mineralogicheskii Zhurnal* **1979**, *13*, 73-78.
19. Reddy, K.M.; Jacob, A.S.; Reddy, B.J. EPR and optical spectra of $Cu^{2+}$ in dioptase. *Ferroelectrics Letters* **1986**, *6*, 103-112.
20. Breuer K.H.; Eysel, W. Structural and chemical varieties of dioptase, $Cu_6[Si_6O_{18}]\cdot 6H_2O$. *Zeitschrift für Kristallographie* **1988**, *184,* 1-11.
21. Huang. Y.P.; Jiang, M.; Wang, L.J.; Feng, W.L. Theoretical investigation of the optical spectra and g factors for $Cu^{2+}$ in dioptase. *Philosophical Magazine* **2008**, *88*, 1701-1704.
22. Wolfram, H. Zur Kristallchemie und Kristallphysik niedrigdimensionales Silicate, Germanate und Arsenate des Kupfers. *Dissertation* **2004**, TU Clausthal.
23. Otto, H.H.; Wolfram, H. New cost-efficient ambient pressure synthesis, Rietfeld analysis and UV-VIS spectrum of Litidionite, $CuNaKSi_4O_{10}$, a weathering-proof ancient pigmente. *Phys. Chem. Miner.* **2017**, to be published.
24. Rudko, G.Yu.; Long, V.C.; Musfeldt, J.L.; Koo, H.J.; Whangbo, M.H.; Revcolevschi, A.; Dhalenne, G.; Bernholdt, D.E. Electronic Transition in Doped and Undoped Copper Germanate. *Chem. Mater.* **2001**, *13,*

939-944.
25. Jahn, H.A.; Teller, E. Stability of Polyatomic Molecules in Degenerate Electronic States. I. Orbital Degeneracy. *Proceedings of the Royal Society A* **1937**, *161*, 220-235.
26. Siemons, W.; Koster, G.; Blank, D.H.A.; Hammond, R.H.; Geballe, T.H.; Beasley, M.R. Tetragonal CuO: A new end member of the 3d transition metal monoxides. *Arxiv* **2008**: 0810.5231v [cond-mat.mtrl-sci].
27. Gerloch, M.; Slade, R. Ligand-field parameters. Cambridge University Press **1973**, Cambridge.
28. Haverkort, M.W. Spin and orbital degrees of freedom in the transition metal oxides and oxide thin films studied by soft x-ray absorption spectroscopy. *Doctoral Thesis* **2005**, University of Köln.
29. Lebernegg, S.; Amthauer, G.; Grodzicki, M. The d-Hamiltonian – A new approach for evaluating optical spectra of transition metal complexes. *Journal of Molecular Structure* **2009**, *924-926*, 473-476
30. Eby, R.K.; Hawthorne, F.C. Structural Relation in Copper Oxysalt Minerals. I. Structural Hierachy. *Acta Crystallographica B* **1993**, *49*, 28-56.
31. Moretti Sala, M.; Bisogni, V.; Aruta, C.; Balestrino, G.; Berger, H.; Brookes, N.B.; DeLuca, G.M.; Castro, D.D.; Grioni, M.; Guarise, M.; Medaglia, P.G.; Miletto Granozio, F.; Minola, M.; Perna, P.; Radovic, M.; Sallustro, M.; Schmitt, T.; Zhou, K.J.; Braikovic, L.; Ghiringhelli, G. Energy and symmetry of dd excitations in undoped layered cuprates measured by Cu $L_3$ resonant inelastic x-ray scattering. *New Journal of Physics* **2011**, *13*, 1-25.
32. Rocquefelte, X.; Schwarz, K.; Blaha, P. Theoretical Investigation of the Magnetic Exchange Interaction in Copper(II) Oxides under Chemical and Physical Pressures. *Nature.com Scientific Reports* **2012**, Article No. 759, 1-7.
33. Otto, H.H. Modeling of a Cubic Antiferromagnetic Cuprate Super-Cage. *World Journal of Condensed Matter Physics* **2015**, *5*, 160-178.
34. Brown, I.D.; Shannon, R.D. Empirical bond-strength - bond-length curves for oxides. *Acta Crystallographica A* **1973,** *29*, 266-282.
35. Otto, H.H. Turbo-Basic program *Valence*, **1980**, University of Regensburg
36. Yamada, M.; Nishi, M.; Akimitsu, J. Electron paramagnetic resonance governed by the Dzyaloshinsky-Moriya antisymmetric exchange interaction in $CuGeO_3$. *Journal of Physics of Condensed Matter* **1996**, *8*, 2625-2640.
37. Hidaka, M.; Hatae, M.; Yamada, I.; Nishi, M.; Akimitsu, J. Re-examination of the room temperature crystal structure of $CuGeO_3$ by x-ray diffraction experiments: observation of new superlattice reflections. *Journal of Physics of Condensed Matter* **1997**, *9*, 809-824.
38. Gippius, A.A.; Morozova, E.N.; Khozeev, D.F.; Vasil'ev, A.N.; Baenitz, M.; Dhalenne, G.; Revcolevschi, A. Non-equivalence of Cu crystal sites in $CuGeO_3$ as evidenced by NQR. *Journal of Physics of Condensed Matter* **2000**, *12*, L71-L75.
39. Li, X.; Kanai, M.; Kawai, T.; Kawai, S. Epitaxial Growth and Properties of $Ca_{1-x}Sr_xCuO_2$ Thin Film (x = 0.18 to 1.0) Prepared by Co-Deposition and Atomic Layer Stacking. *Japanese Journal of Applied Physics* **1992**, *31*, L217-L220.
40. Giester, G. and Rieck, B. Effenbergite, a New Mineral from the Kalahari manganese field, South Africa; description and crystal structure. *Mineralogical Magazine* **1994**, *58*, 663-670.
41. Chakoumakos, B.C.; Fernandez-Baca, J.A.; Boatner, L.A. Refinement of the Structures of the Layer Silicates $MCuSi_4O_{10}$ (M = Ca,Sr,Ba) by Rietfeld Analysis of Neutron Powder Diffraction Data. *Journal of Solide State Chemistry* **1993**, *103*, 105-113.
42. Steinberg, H.; Meibohm, M.; Hofmann, W.; Otto, H.H. Neue Synthesemethode und Rietfeld-Verfeinerung von $CaCuSi_4O_{10}$. *Beihefte zu European Journal of Mineralogy* **1999**, *11*, 219.
43. Yoshisa, A.; Bagum, R.; Iguchi, Y.; Okube, M.; Mashimo, T. Structure of $Y_2BaCuO_5$ synthesized under strong gravity field. *Photon Factory Activity Report* **2010** #28 Part B (2011).
Sato, S.; Nakada, J. Structure of $Y_2BaCuO_5$: a refinement by single crystal X-ray diffraction. *Acta Crystallographica C* **1989**, *45*, 523-525.
44. Ribbe, P.H.; Gibbs, G.V.; Hamil, M.M. A refinement of the structure of dioptase, $Cu_6[Si_6O_{18}]\cdot 6H_2O$. *American Mineralogist* **1977**, *62*, 807-811.
45. Belokoneva, E.L.; Gubina, Y.K.; Forsyth, J.B. The charge-density distribution and antiferromagnetic properties of azurite $Cu_3[CO_3]_2(OH)_2$. *Physics and Chemistry of Minerals* **2001**, *28*, 498-507.
46. Bacon, G.E.; Titterton, D.H. Neutron-diffraction studies of $CuSO_4\cdot 5H_2O$ and $CuSO_4\cdot 5D_2O$. *Zeitschrift für*

*Kristallographie* **1975**, *141*, 330-341.
47. Hawthorne, F.C. Lammerite, $Cu_3(AsO_4)_2$, a modulated close-packed structure. *American Mineralogist* **1986**, *71*, 206-209.
48. Henderson, R. R.; Yang, H.; Downs, R.T.; Jenkins, R.A. Redetermination of conichalcite, $CaCu(AsO_4)(OH)$. *Acta Crystallographica E* **2008**, *64*, i53-i54,1-7.
49. Häflinger, P.S.; Gerber, S.; Pramod, R.; Schnells, V.I.; dalla Plazza, B.; Chati, R.; Pomjakushin, V.; Conder, K.; Pomjakushina, E.; Le Dreau, L.; Christensen, N.B.; Syljuasen, O.F.; Normand, B.; Rønnow, H.M. Quantum and thermal motion, oxygen isotope effect, and superexchange distribution in $La_2CuO_4$. *Physical Review B* **2014**, *89*, 085113, 1-13
50. Miller, L.L.; Wang, X.L.; Wang, S.X.; Stassis, C.; Johnston, D.C.; Faber, J.; Loong, C.K. Synthesis, structure, and properties of $Sr_2CuO_2Cl_2$. *Physical Review B* **1990,** *41*, 1921-1925.
51. Reddy, R.R.; Reddy, S.L.; Rao, P.S.; Frost, R.L. Optical absorption an EPR studies on tenorite mineral. *Spectrochimica Acta A* **2010**, *75*, 28-31.
52. Reddy, B.J.; Frost, R.L.; Martens, W.N. Characterization of Conichalcite by SEM, FTIR, Raman and electronic reflectance spectroscopy. *Mineralogical Magazine* **2005**, *69*, 155-167.
53. Baran, E.J.; Cicileo, G.P. The electronic spectrum of $Y_2BaCuO_5$. *Journal of Material Science Letters* **1990**, *9*, 1-2.
54. Reddy, B.J.; Sarma, K.B.N. Absorption Spectra of $Cu^{2+}$ in Azurite. *Solid State Communications* **1981**, *38*, 547-549.
55. Redhammer, G.J.; Koll, L.; Bernroider, M.; Tippelt, G.; Amthauer, G.; Roth, G. $Co^{2+}$-$Cu^{2+}$ substitution in bieberite solid-solution series, $(Co_{1-x}Cu_x)SO_4{\cdot}7H_2O$, $0.00 \leq x \leq 0.46$: Synthesis, single-crystal structure analysis, and optical spectroscopy. *American Mineralogist* **2007**, ***192***, 532-545.
56. Kendrick, E.; Kirk, C.J.; Dann, S.E. Structure and colour properties in the Egyptian Blue Family, $M_{1-x}$ $M'_x CuSi_4O_{10}$ as a function of M, M'where M, M' = Ca, Sr and Ba. *Dyes and Pigments* **2007**, *73*, 13-18.
57. Ford, R.J. and Hitchman, M.A. Single crystal electronic and EPR spectra of $CaCuSi_4O_{10}$, a synthetic silicate containing copper(II) in a four-coordinate, planar ligand. *Inorganica Chimica Acta* **1979**, *33*, L167-L170.
58. The NIST Reference on Constants, Units, and Uncertainty. *NIST*, Gaithersburg, MD20899, USA.

**Supplemented material**

**Table 7.** Comparison of scaling factors used for Cuprates in comparison to dioptase group compounds.

| Scaling factor notation | Cuprates | Dioptase group | Ratio |
|---|---|---|---|
| $Q_e^2$ | 1.755 | 1.733 | 1.013 |
| $Q_a^2$ | 0.936 | 0.761 | 1.230 |
| $\eta_s$ | 1.587 | 1.276 | 1.244 |
| $\eta_{te}$ | 2.180 | 1.738 | 1.254 |